# Diffuse Neutron Scattering Study of Magnetic Correlations in half-doped $La_{0.5}Ca_{0.5-x}Sr_xMnO_3$ (x = 0.1, 0.3 and 0.4) Manganites

I. Dhiman[1], A. Das[1,*], R. Mittal[1], Y. Su[2], A. Kumar[1], and A. Radulescu[2]

[1]Solid State Physics Division, Bhabha Atomic Research Centre, Mumbai - 400085, India

[2]Juelich Centre for Neutron Science, IFF, Forschungszentrum Juelich, Outstation at FRM II, Lichtenbergstrasse 1, D-85747 Garching, Germany.

## Abstract

The short range ordered magnetic correlations have been studied in half doped $La_{0.5}Ca_{0.5-x}Sr_xMnO_3$ (x = 0.1, 0.3 and 0.4) compounds by polarized neutron scattering technique. On doping $Sr^{2+}$ for $Ca^{2+}$ ion, these compounds with x = 0.1, 0.3, and 0.4 exhibit CE-type, mixture of CE-type and A-type, and A-type antiferromagnetic ordering, respectively. Magnetic diffuse scattering is observed in all the compounds above and below their respective magnetic ordering temperatures and is attributed to magnetic polarons. The correlations are primarily ferromagnetic in nature above $T_N$, although a small antiferromagnetic contribution is also evident. Additionally, in samples x = 0.1 and 0.3 with CE-type antiferromagnetic ordering, superlattice diffuse reflections are observed indicating correlations between magnetic polarons. On lowering temperature below $T_N$ the diffuse scattering corresponding to ferromagnetic correlations is suppressed and the long range ordered antiferromagnetic state is established. However, the short range ordered correlations indicated by enhanced spin flip scattering at low Q coexist with long range ordered state down to 3K. In x = 0.4 sample with A-type antiferromagnetic ordering, superlattice diffuse reflections are absent. Additionally, in comparison to x = 0.1 and 0.3 sample, the enhanced spin flip scattering at low Q is reduced at 310K, and as temperature is reduced below 200K, it becomes negligibly low. The variation of radial correlation function, g(r) with temperature indicates rapid suppression of ferromagnetic correlations at the first nearest neighbor on approaching $T_N$. Sample x = 0.4 exhibits growth of ferromagnetic phase at intermediate temperatures (~ 200K). This has been further explored using SANS and neutron depolarization techniques.

*E mail: adas@barc.gov.in

# I. INTRODUCTION

The complex interplay between charge, spin, orbital and lattice degree of freedom is responsible for the rich phase diagram in doped perovskite $La_{1-x}Ca_xMnO_3$ manganites.[1,2] Neutron scattering studies in doping regime $0.15 \leq x \leq 0.30$ have shown that the transport properties of these materials are controlled by the competition between short range charge correlations and long range ferromagnetic double exchange interactions.[3,4] Theoretical understanding enunciates that double exchange interaction alone is insufficient and strong electron – phonon coupling is also required. Origin of this coupling is proposed to be due to lattice polarons and dynamic Jahn-Teller distortions.[5-8] Neutron and x – ray scattering experiments are directly sensitive to both polarons and their correlations and therefore, they make an important contribution to studies of polarons.

In manganites, strong electron – phonon coupling results in the formation of localized charge carriers associated with lattice distortions (polarons) in the paramagnetic insulating regime. Early evidence of polarons has been obtained from transport studies.[9,10] For $La_{0.7}Ca_{0.3}MnO_3$, these polarons take the form of correlations with an ordering of wave vector ≈ (¼, ¼, 0).[11] This type of ordering becomes long range at half doping, having CE-type (charge exchange) antiferromagnetic, charge and orbitally ordered state with equal number of $Mn^{3+}$ and $Mn^{4+}$ ions.[12,13] In compounds with $x < ½$, the CE-type antiferromagnetic structure is frustrated and is observed in the insulating state of manganite in the form of nanoscale structural correlations.[14-17] The onset of ferromagnetism below $T_C$ (~ 257K), leads to melting of these CE-type correlations observed in the insulating regime just above $T_C$. As a result no diffuse scattering is observed far below $T_C$.[15-20] Previous x – ray and neutron scattering study on half doped perovskite manganite $(Nd_{0.125}Sm_{0.875})_{0.52}Sr_{0.48}MnO_3$ and the layered manganite $La_{1.2}Sr_{1.8}Mn_2O_7$ also revealed a direct evidence for the formation of lattice polarons.[21,22] Mellergård et al. have reported the absence of local lattice distortions (lattice polarons) below $T_C$ for $La_{1-x}Sr_xMnO_3$ ($x = 0.2$ and $0.4$) compounds using neutron diffraction and Reverse Monte Carlo analysis.[23] They linked the observed distortions above $T_C$ with $Mn^{4+}$ ion, which is different from the Jahn-Teller type associated with the $Mn^{3+}$ ion. Also, the magnetic-moment pair correlation function calculation gives evidence for short range magnetic correlations (magnetic polarons). These magnetic polarons are correlated with the local lattice distortions (lattice polarons). The ferromagnetic correlation is associated with shorter Mn-Mn distances and antiferromagnetic correlation with the longer distance.

In this report we study the short range magnetic correlations in $La_{0.5}Ca_{0.5-x}Sr_xMnO_3$ (x = 0.1, 0.3 and 0.4) series above and below the magnetic ordering temperature by polarized neutron scattering technique. In the parent compound (x = 0), as a result of 1:1 ratio of $Mn^{3+}$ and $Mn^{4+}$ ions, the charge and orbitally ordered CE-type state is found to be most stable. It exhibits ferromagnetic transition at $T_C \approx 230K$ and a charge and orbitally ordered antiferromagnetic insulating transition at $T_N \approx 170K$.[24] In the previously reported neutron diffraction study using unpolarized neutrons on $La_{0.5}Ca_{0.5-x}Sr_xMnO_3$ ($0.1 \leq x \leq 0.5$) compounds we have shown the suppression of CE-type antiferromagnetic phase with progressive increase in Sr doping and establishment of ferromagnetic phase.[25] The CE-type antiferromagnetic phase is observed for x = 0.1, 0.2 and 0.3 samples with antiferromagnetic transition temperatures 150K, 100K and 75K, respectively. At x = 0.4, the CE-type antiferromagnetic structure is fully suppressed and A-type antiferromagnetic phase is observed with the transition temperature $T_N \approx 200K$. For x = 0.5 sample, the A-type antiferromagnetic transition temperature is reduced to 125K and the long range ferromagnetically ordered phase is established at all temperatures below 310K. Using polarization analysis techniques we are able to separate the magnetic diffuse scattering from other contributions such as nuclear and thermal diffuse scattering. As a result we provide a clear evidence for the presence of magnetic diffuse scattering coexisting with long range ordered CE-type and A-type antiferromagnetic phase. Additionally, small angle neutron scattering (SANS) and neutron depolarization measurements have been carried out for sample x = 0.4 which exhibits ferromagnetic phase in the intermediate temperature regime.

## II. EXPERIMENT

The polycrystalline samples $La_{0.5}Ca_{0.5-x}Sr_xMnO_3$ (x = 0.1, 0.3 and 0.4) were synthesized by conventional solid-state reaction method reported elsewhere. The phase purity of all the samples is confirmed by x ray and neutron diffraction techniques reported previously.[25] Polarized neutron diffraction ($\lambda$ = 4.74Å) measurements in the angular range $20° \leq 2\theta \leq 125°$ were carried out on the diffuse neutron scattering (DNS) spectrometer at FRM-II reactor, at several temperatures between 3K and 310K. Normal collimators and Beryllium filter for removing $\lambda/2$ contamination were used in the course of the experiment. We have carried out xyz polarization analyses to separate the magnetic scattering from nuclear and spin incoherent scattering. SANS measurements ($\lambda$ = 10Å) as a function of temperature (20K $\leq$ T $\leq$ 300K) in zero magnetic field for Q range between $10^{-3}$ $Å^{-1}$ and $0.30Å^{-1}$ was carried out on SANS instrument (KWS-2) at FRM II reactor. The position-sensitive (two-dimensional) Anger – type scintillation detector (60 × 60 $cm^2$ 6Li glass scintillator 1mm thick and an array of 8 × 8 photomultipliers) with a resolution of 0.5 × 0.5 $cm^2$ were used to carry out SANS measurements. The 2-D raw data were corrected for the scattering from empty can and cryostat windows, the electronic and background noise and calibrated to absolute scale using a plexiglass standard sample. After azimuthal integration of the 2-D data the scattered intensity I(Q) as a function of the scattering vector Q was obtained. Neutron depolarization measurements ($\lambda$ = 1.205Å) were carried out on the polarized neutron spectrometer at Dhruva reactor, Bhabha Atomic Research Centre, Mumbai, India, with $Cu_2MnAl$ (1 1 1) as polarizer and $Co_2Fe$ (2 0 0) as analyzer.

## III. RESULTS AND DISCUSSION

### A. Polarized Neutron Scattering

The samples studied in the series $La_{0.5}Ca_{0.5-x}Sr_xMnO_3$ ($0 < x \leq 0.3$) crystallize with orthorhombic structure (space group *Pnma*) and x = 0.4 crystallizes with two orthorhombic phases in the space group *Pnma* and *Fmmm*. The structural and magnetic properties of these samples have been reported previously.[25] In this study, three samples have been chosen which exhibit distinct magnetic structures. The x = 0.1 compound exhibits CE-type antiferromagnetic spin structure, x = 0.3 exhibits mixture of CE-type and A-type antiferromagnetic structure while, x = 0.4 undergoes A-type antiferromagnetic ordering at low temperatures.[25] The xyz polarization analysis allows us to separate the nuclear and magnetic contributions. All the spin-flip (SF) scattering is purely magnetic in nature. The non-spin-flip (NSF) scattering does not contain magnetic scattering component and has contribution only from nuclear coherent and isotopic incoherent scattering.[26,27] Figure 1 shows a typical diffraction pattern with separated SF and NSF contributions in x = 0.1 sample at 310 K. The NSF scattering indicating the nuclear Bragg reflections (1 0 1) (0 2 0) at $2\theta \sim 75.5°$ and (2 0 0) (0 0 2) (1 2 1) at $2\theta \sim 123°$ is in agreement with the previously reported structural studies on this compound.[25]

The SF and NSF scattering for x = 0.1 sample at 310K is shown in figure 2(a). At this temperature, in the paramagnetic region, magnetic diffuse scattering peak is primarily centered at $2\theta \sim 75.5°$ and a weak superlattice diffuse scattering peak is observed at $\sim 33.5°$, in addition to enhanced scattering at low $2\theta$ values in SF scattering. The broad diffuse scattering peak at $2\theta \sim 75.5°$ is centered around fundamental Bragg reflection (020) (101) (observed for NSF scattering). This peak corresponds to short range ferromagnetic correlations, as it is observed around the fundamental Bragg reflection indexed as (1 0 1) (0 2 0). At 310K below $2\theta \sim 30°$, the SF component also shows an enhanced scattering. Existence of enhanced scattering in the SF and NSF scattering components have been attributed to the existence of magnetic and lattice polarons, respectively.[28] However, in the present study enhanced scattering is observed only in the SF component, indicating the existence of magnetic polarons alone. Similarly, the superlattice reflection is attributed to arise from polaron-polaron correlation. The enhanced scattering at low Q is fitted to a lorentzian-type Q dependence, $I = I_0 / [(1/\xi)^2 + Q^2]$, where $\xi$ is the correlation length (Ornstein-Zernike form), as shown in figure 2(a). The obtained correlation length at 310K for x = 0.1 sample is ~ 3.4(9) Å, which is of the order of Mn-Mn distance. The diffuse scattering peak evident at $2\theta \sim 33.5°$ indicated by an arrow in figure 2(a) is a superlattice peak. The superlattice diffuse peak at $2\theta \sim 33.5°$ is very weak at 310K. On lowering temperature below 310K, the superlattice diffuse peak at $2\theta \sim 33.5°$ becomes more pronounced. For x = 0.1 sample the SF scattering at 175Kis shown in figure 2(b). However, the broadness of the peak suggests the short range nature of antiferromagnetic correlations. It is of interest to note that these short range antiferromagnetic and ferromagnetic correlations are observed at 310K, which is much above the transition temperature ($T_N \sim 150$K and $T_C \sim 244$K) for this compound. Below 175K, on approaching $T_N$, the short range ordered antiferromagnetic correlations are suppressed with the onset of long range CE-type antiferromagnetic ordering. Figure 2(c) displays the SF scattering at 3K indicating the long range ordered antiferromagnetic superlattice reflections indexed to a CE-type antiferromagnetic structure reported previously in this compound. Additionally, the absence of SF scattering in the fundamental nuclear reflections rules out the possibility of long-range ferromagnetic ordering coexisting with antiferromagnetic ordering in this sample. The coexistence of the short range ferromagnetic and antiferromagnetic correlations at 310K indicates the presence competing magnetic interactions. The temperature dependence of the integrated intensity obtained using a lorentzian peak shape function fit to the ferromagnetic ($2\theta \sim 33.5°$) and antiferromagnetic ($2\theta \sim 75.5°$) diffuse scattering peaks for sample x = 0.1, are shown in figure 3. The ferromagnetic diffuse scattering intensity shows a maximum at ~ 225K. This maximum is close to the ferromagnetic transition temperature ($T_C \sim$ 244K), obtained from previously reported M(T) measurements.[25] On lowering of temperature below 225K, the short range ordered ferromagnetic correlations are suppressed while the antiferromagnetic correlations continue to increase. The onset of long range antiferromagnetic ordering below 175K coincides with rapid suppression of short range ferromagnetic ordering.

The SF scattering shown in figure 4(a) for $La_{0.5}Ca_{0.2}Sr_{0.3}MnO_3$ (x = 0.3) compound displays a superlattice peak centered at $2\theta \sim 33.5°$ and a peak at $2\theta \sim 75.5°$, centered around the fundamental nuclear reflections (1 0 1) (0 2 0), similar to x = 0.1 compound. These two peaks correspond to antiferromagnetic and ferromagnetic short range ordered correlations, respectively as described for x = 0.1 sample. At 275K the intensity of superlattice diffuse peak at $2\theta \sim 33.5°$ increases in comparison with 310K, as evident in figure 4(b). Below 150K, these short range ordered antiferromagnetic correlations are suppressed with the onset of long range ordering. Figure 4(c) displays the SF scattering for x = 0.3 sample at 3K. The superlattice reflections corresponding to long range ordered CE-type antiferromagnetic spin structure are observed, in concurrence with our previously reported neutron diffraction study.[25] In addition, at low temperature few additional superlattice reflections (at 2θ ~ 37° and ~ 88°) other than the ones corresponding to CE-type antiferromagnetic spin structure are observed. This antiferromagnetic phase is identified as having an A-type spin structure. We failed to detect this phase in our previous neutron diffraction studies.[25] Therefore, at low temperature the magnetic phase of this compound consists of a mixture of CE-type and A-type antiferromagnetic state. Both the magnetic phases have identical transition temperatures ($T_N$). The short range ordered ferromagnetic interactions are similar to x = 0.1 sample and no long range ferromagnetic ordering is observed down to 3K.

The SF scattering at 310K for x = 0.4 sample displayed in figure 5(a) provides evidence of diffuse ferromagnetic correlations and enhanced scattering at low 2θ values. This sample with A-type antiferromagnetic ordering does not display the superlattice diffuse reflection at 2θ ~ 33.5°, which is observed in x = 0.1 and 0.3 compounds having dominant CE-type antiferromagnetic spin structure. The diffuse ferromagnetic scattering peak at 2θ ~ 75.5°, observed in figure 5(a) and (b) is similar to x = 0.1 and 0.3 compounds. Figure 5(c) exhibits the SF scattering (purely magnetic scattering) at 3K. Below 250K well defined superlattice Bragg reflections are observed indicating the onset of long-range ordered A-type antiferromagnetic structure. The superlattice reflections corresponding to long range ordered A-type antiferromagnetic phase are evident in figure 5(c). Integrated intensity of the ferromagnetic diffuse peak centered at 2θ ~ 75.5° is shown in the inset of figure 5(c). This peak exhibits a maximum at 200K, with the establishment of long range ordered ferromagnetic interactions between 250K - 150K. Below 150K, this ferromagnetic phase is suppressed. This behavior is in agreement with the maximum in magnetization reported earlier and minimum in neutron depolarization (figure 10) measurements discussed subsequently. Also, our neutron diffraction measurements reported earlier display similar maximum in integrated intensity versus temperature plot for (1 0 1) (0 2 0) reflection.[25]

From the full width at half maximum (ΔQ) of the diffuse scattering centered at $2\theta \sim 75.5°$, the size of the short range ordered regions is estimated. ΔQ for the ferromagnetic diffuse scattering peak was estimated by fitting it to a lorentzian peak shape function. The lorentzian fit for x = 0.1, 0.3 and 0.4 compounds is shown in figure 2(a), 4(a) and 5(a), respectively. The correlation length, $\xi$ (= $2\pi/\Delta Q$) at 310K has values of 13(2) Å, 10(1) Å and 15(2) Å for x = 0.1, 0.3 and x = 0.4 samples, respectively. No significant change in $\xi$ is observed with variation of temperature.

In figure 6 the temperature dependence of magnetic diffuse scattering intensity at Q = 0.46Å$^{-1}$ ($2\theta \approx 20°$) for $La_{0.5}Ca_{0.5-x}Sr_xMnO_3$ series with x = 0.1, 0.3 and 0.4 is displayed. For x = 0.1 and 0.3 samples, similar temperature dependence of magnetic diffuse scattering intensity is observed. The short range ordered antiferromagnetic correlation in x = 0.1 and 0.3 compounds coexist with the long range CE-type antiferromagnetic ordering at the lowest temperature of 3K. However, for x = 0.4 composition these short range ordered antiferromagnetic correlations are much reduced as compared to x = 0.1 and 0.3 compounds.

The evidence of diffuse scattering in polarization analysis measurements of these samples indicates the existence of magnetic polarons above the transition temperature. Absence of diffuse scattering in the nuclear coherent scattering (NSF scattering) component, indicate the absence of lattice polarons. Similar study using neutron polarization analysis technique on half doped $Nd_{0.5}Pb_{0.5}MnO_3$ compound has been reported by Clausen et al.[29] Strong diffuse magnetic scattering is observed above $T_C$, attributed to magnetic polarons (SF scattering) while lattice polarons (NSF scattering) are not observed. Therefore, short range ordered antiferromagnetic correlation are visible above the transition temperature and are precursors to CE-type antiferromagnetic phase. This is distinct from similar studies reported in ferromagnetic compound $La_{0.7}Ca_{0.3}MnO_3$ where short range ordered CE-type polarons are found to exist above $T_C$ in the insulating state.[14]

Mellergård et al.[23] have associated the ferromagnetic correlations with the shorter Mn–Mn distances and antiferromagnetic correlation with longer distances. To obtain a similar behavior on the antiferromagnetic and ferromagnetic correlations as a function of distance we have calculated the radial correlation function, g(r). The g(r) is obtained from the SF scattering data, using the following expression[30],

$$g(r) = \int_{Q_l}^{Q_h} I_{mag}(Q) f(Q)^{-2} Q \sin(Qr) dQ \quad (1)$$

where, Q (= $4\pi \mathrm{Sin}\theta/\lambda$) is the scattering vector, $I_{mag}(Q)$ is the magnetic scattering intensity, and f(Q) is the magnetic scattering form factor. Assuming that the interactions are isotropic, g(r) is related to spin-spin correlation function by the following expression[31],

$$g(r) = \frac{1}{S(S+1)} \sum_{r'} \langle S_0 \cdot S_{r'} \rangle \, \delta(|r| - |r'|) \quad (2)$$

This expression is a sum of spin-spin correlation function at distance r. Figures 7, 8 and 9 show the variation of g(r) calculated for various temperatures for sample x = 0.1, 0.3 and 0.4, respectively. It is apparent from these figures that there is a strong change in the average character of the short-range magnetic interactions with nearest neighbors, when approaching the transition temperature from the higher temperature. Figure 7 shows the radial correlation function g(r) at 310K, 200K, 175K, and 125K for $La_{0.5}Ca_{0.4}Sr_{0.1}MnO_3$ (x = 0.1) sample, with first to fifth nearest neighbor Mn pairs indicated by arrow. Well above the ferromagnetic transition temperature, the Mn-Mn pairs which are nearest neighbors (~4Å) build up ferromagnetic correlations within the paramagnetic matrix, as can be inferred from the positive value of g(r) for small r. This behavior is also evident in figure 2, where diffuse scattering is observed at 2θ ~ 75.5°, indicative of short range ordered ferromagnetic correlations. In addition, for Mn-Mn pairs located at next-nearest neighbor distances ~ 5.4Å, g(r) shows negative value, as shown in figure 7. This suggests the existence of magnetic correlations of an antiferromagnetic character. The temperature evolution of these correlations for x = 0.1 sample, show that the ferromagnetic correlations at the first nearest neighbor distance are strongly suppressed as temperature is reduced. On the other hand, the antiferromagnetic correlations do not exhibit appreciable change with temperature. Figure 8 displays g(r) for x = 0.3 sample at 310K and 150K. Arrows indicate the first two nearest neighbor bond distance for Mn pairs. Similar to x = 0.1 sample, the positive value of g(r) for nearest neighbor Mn – Mn pairs and the negative value of g(r) for next nearest neighbor Mn – Mn pairs indicate the existence of ferromagnetic and antiferromagnetic magnetic correlations, respectively. Figure 9 shows the radial correlation function g(r) for $La_{0.5}Ca_{0.1}Sr_{0.4}MnO_3$ (x = 0.4) sample at 275K and 150K. The three nearest neighbor bond distances for Mn pairs are indicated by arrows. In this sample g(r) has a positive value for Mn pairs separated by distances ~ 3.8Å. This indicates that nearest neighbor correlations are ferromagnetic in nature. On reducing temperature the ferromagnetic correlations does not change appreciably. The g(r) for the next nearest neighbor Mn – Mn pairs (~ 6Å) is negative, indicating the antiferromagnetic nature of the correlations. On reducing temperature below $T_N$, the antiferromagnetic correlations are strongly enhanced while the ferromagnetic correlations at the first nearest neighbor distance are only moderately influenced. This behavior may be correlated with the A-type antiferromagnetic ordering observed in this compound, where antiferromagnetic coupling exists between ferromagnetic planes.

## B. Neutron Depolarization

Unlike, x = 0.1 and 0.3 compounds, where no evidence of long range ferromagnetic ordering is observed, sample x = 0.4 exhibits an unusual behavior of increase in ferromagnetic behavior between 150K and 250K (as shown by the increase in SF intensity of the fundamental Bragg reflections (101) (020) in the inset of figure 5c). This behavior is further studied using neutron depolarization. Neutron depolarization is a technique suitable for the detection of magnetic inhomogeneities on mesoscopic length scale ranging from 1000 Å to several microns. In the present study, we have measured flipping ratio R (ratio of the transmitted intensities for two spin states of the incident neutron spin) which is a measure of the transmitted beam polarization. R is expressed in the form[32,33]

$$R = \frac{1-P_i D P_A}{1+(2f-1)P_i D P_A}$$

where, $P_i$ is the incident beam polarization, $P_A$ is the efficiency of the analyzer crystal, f is the rf flipper efficiency and D is the depolarization coefficient. In the absence of any depolarization in sample, D = 1. $P_iD$ is thus the transmitted beam polarization.

Figure 10 shows the temperature dependence of transmitted neutron beam polarization (P) for sample $La_{0.5}Ca_{0.1}Sr_{0.4}MnO_3$ (x = 0.4), with H = 50Oe, under zero-field-cooled conditions. For this sample, polarization remains constant up to ~232K. Below 232K, polarization decreases rapidly, reaching a minimum at ~ 180K. As temperature is reduced further, polarization again increases almost reaching the same value as in the paramagnetic state. The decrease in polarization below 232K indicates the onset of ferromagnetic ordering. This behavior correlates well with our previously reported magnetization M(T) and neutron diffraction study on this compound.[25] The rapid suppression of the depolarization signal below 180K indicates the reduction of domain size (δ) and / or domain magnetization (B). This behavior correlates with the observation of a maximum at ~ 200K in integrated intensity of SF scattering peak (1 0 1) (0 2 0) for this sample, shown in inset of figure 5(c). The reduction of ferromagnetic nature in antiferromagnetic phase (with A – type spin structure) below 180K is a consequence of competing interactions between antiferromagnetic and ferromagnetic interactions.

An estimate of domain size in the ferromagnetic region is obtained using the expression

$$P_f = P_i \exp(-\alpha(d/\delta))\langle\phi_\delta\rangle^2$$

where, $P_f$ and $P_i$ are the transmitted beam and incident beam polarization, respectively, $\alpha$ is a dimensionless parameter = 1/3, d is the sample thickness, $\delta$ is a typical domain length and the precession angle $\phi_\delta = (4.63 \times 10^{-10}\ \mathrm{Oe}^{-1}\ \mathrm{Å}^{-2})\ \lambda\delta B$.[34,35] The domain magnetization, B is obtained from the bulk magnetization. This expression is valid in the limit where domains are randomly oriented and the Larmor phase of neutron spin due to the internal magnetic field of sample < 2π over a typical domain length scale. Our measurements were carried out in low field far away from the saturation field and therefore satisfy the assumption of this model. The estimated domain size in the present sample at T = 2K is ~ 0.8μm.

The depolarization measurements were also performed for sample x = 0.1, and 0.3. However, no change in transmitted beam polarization was observed down to lowest temperature of 2K. This is in agreement with the SF scattering measurements on these samples where no evidence of enhancement in the intensity of the fundamental Bragg reflections (101) (020) is observed. Thus we rule out presence of ferromagnetic correlations in the CE-type antiferromagnetic phase of samples x = 0.1 and 0.3.

### C. Small Angle Neutron Scattering

The evolution of magnetic scattering intensity [I(20) - I(300)] as a function of Q (0.0073 $\mathrm{Å}^{-1} \leq Q \leq 0.080\mathrm{Å}^{-1}$) for $La_{0.5}Ca_{0.1}Sr_{0.4}MnO_3$ (x = 0.4) compound is shown in figure 11. The Q range in which measurements were carried out corresponds to length scale of 40Å – 1000Å. This figure [I(T) - I(300)] is representative for all the temperatures in the range 20K ≤ T < 300K, To estimate the Q dependence of the magnetic scattering in SANS, intensity at each temperature was subtracted from data at 300K, taken as the background intensity. The pure magnetic scattering intensity [I(20K) - I(300K)] thus obtained, in the Q range 0.007 – 0.08Åis best described by squared Lorentzian type function, $I = I_0 / [(1/\xi)^2 + Q^2]^2$, where $\xi$ is the spin-spin correlation length and $I_0$ is Lorentz amplitude related to the bulk susceptibility. The temperature dependence of fitting parameters is shown in the inset of figure 11. The correlation length varies from 123(4) Å at 20K to 153(3) Å at 250K, exhibiting a maximum at 200K having a value of 232(12) Å. The ferromagnetic correlation length obtained from SANS measurements is much smaller than the size of the domains obtained from depolarization measurements due to the difference in length scales at which the two techniques probe. Such differing values from the two measurements have been reported previously.[36,37] The Lorentz amplitude $I_0$ also behaves in a similar manner, with a maximum at 200K, following a behavior similar to M(T), reported previously. The squared-Lorentzian behavior in SANS intensity is expected for static cluster scattering. Debye et al. reported from theoretical study that a squared Lorentzian type function would be observed for an array of random shapes, sizes and distribution in a solid matrix.[38,39] In our analysis, additional Lorentzian term which describes a critical scattering was not necessary to fit the data.[40]

Figure 12 shows the temperature dependence of the scattered neutron intensity at Q value 0.01Å$^{-1}$ for sample $La_{0.5}Ca_{0.1}Sr_{0.4}MnO_3$ (x = 0.4). At a length scales, of ~ 63nm (Q = 0.01Å$^{-1}$), intensity as a function of temperature displays a maximum at 200K. This coincides with the maximum in magnetization reported earlier and minimum in transmitted neutron beam polarization P (figure 10) (~ 180K). This suggests that the nature of this enhancement is magnetic. In our previous neutron diffraction experiments, we observed a maximum in integrated intensity versus temperature plot for (1 0 1) (0 2 0) nuclear peaks at ~ 200K, indicating the presence of ferromagnetic interactions.

# IV. CONCLUSION

The study of magnetic correlations in $La_{0.5}Ca_{0.5-x}Sr_xMnO_3$ (x = 0.1, 0.3 and 0.4) above and below the ordering temperature is carried out using polarized neutron scattering, neutron depolarization and small-angle neutron scattering techniques. On doping $Sr^{2+}$ ion for $Ca^{2+}$ ion these compounds x = 0.1, 0.3, 0.4 exhibit CE-type, mixture of CE-type and A-type, and A-type antiferromagnetic ordering, respectively. Magnetic diffuse scattering is observed in all the compounds above and below their respective magnetic ordering temperatures and is attributed to magnetic polarons. The correlations are primarily ferromagnetic in nature above $T_N$, although a small antiferromagnetic contribution is also evident. Additionally, in samples x = 0.1 and 0.3 with CE-type antiferromagnetic ordering superlattice diffuse reflections are observed indicating correlations between magnetic polarons. On lowering temperature below $T_N$ the diffuse scattering corresponding to ferromagnetic correlations is suppressed and the long range ordered antiferromagnetic state is established. However, the short range ordered correlations indicated by enhanced spin flip scattering at low Q coexist with long range ordered state down to 3K. In x = 0.4 sample with A-type antiferromagnetic order superlattice diffuse reflections are absent. The enhanced spin flip scattering at low Q is much reduced at 310K, in comparison to x = 0.1 and 0.3 sample. As temperature is decreased below 200K, it becomes negligibly small. The variation of radial correlation function, g(r) with temperature indicates rapid suppression of ferromagnetic correlations at the first nearest neighbor on approaching $T_N$. Sample x = 0.4 exhibits growth of ferromagnetic phase in intermediate temperatures. This has been further explored using SANS and neutron depolarization techniques and allow us to estimate a spin-spin correlation length ($\xi$) of ~ 232(12) Å and a domain size of ~0.8μm at 200K.

## Figure Captions

**Figure 1:** The nuclear (NSF) and magnetic (SF) contributions to the total scattering for $La_{0.5}Ca_{0.4}Sr_{0.1}MnO_3$ (x = 0.1) at 310K from xyz polarization analysis.

**Figure 2:** (a) The spin-flip (SF) scattering for sample $La_{0.5}Ca_{0.4}Sr_{0.1}MnO_3$ (x = 0.1) at 310K is shown on left hand side scale and non-spin flip (NSF) scattering on right hand side scale. Arrow indicates the superlattice magnetic diffuse scattering peak centered at $2\theta \sim 33.5°$. Also shown using continuous line in spin-flip (SF) scattering is the fit to a lorentzian type function described in the text. In (b) SF scattering at 175K and in (c) at 3K is shown.

**Figure 3:** The temperature dependence of integrated intensity obtained using a lorentzian function fit to the antiferromagnetic and ferromagnetic diffuse scattering peaks at $2\theta \sim 33.5°$ and $\sim 75.5°$, respectively for sample $La_{0.5}Ca_{0.4}Sr_{0.1}MnO_3$ (x = 0.1). The continuous lines are a guide for the eye.

**Figure 4:** (a) The spin-flip (SF) scattering for sample $La_{0.5}Ca_{0.2}Sr_{0.3}MnO_3$ (x = 0.3) at 310K is shown on left hand side scale and non-spin flip (NSF) scattering, on right hand side scale, continuous line in spin-flip (SF) scattering is the fit to a lorentzian type function described in the text. Arrow indicates the superlattice diffuse scattering peak centered at $2\theta \sim 33.5°$. In (b) SF scattering at 250K and in (c) at 3K is shown.

**Figure 5:** (a) The spin-flip (SF) scattering for sample $La_{0.5}Ca_{0.1}Sr_{0.4}MnO_3$ (x = 0.4) at 310K is shown on left hand side scale and non-spin flip (NSF) scattering, on right hand side scale, continuous line in spin-flip (SF) scattering shows the fit to a lorentzian function. Arrow indicates the superlattice diffuse scattering peak centered at $2\theta \sim 33.5°$. In (b) SF scattering at 275K and in (c) at 3K is shown. The inset in (c) is the integrated intensity of the diffuse Bragg peak (1 0 1) (0 2 0) at $2\theta \approx 75.5°$.

**Figure 6:** Evolution of neutron scattering intensity as a function of temperature at Q = 0.46 $Å^{-1}$ ($2\theta \approx 20°$) for $La_{0.5}Ca_{0.5-x}Sr_xMnO_3$ series, where x = 0.1, 0.3 and 0.4. The continuous lines are a guide for the eye.

**Figure 7:** Radial correlation function g(r) for $La_{0.5}Ca_{0.4}Sr_{0.1}MnO_3$ (x = 0.1) at 310K, 175K, and 125K. Arrows indicate the five nearest neighbor bond distances for Mn sublattice.

**Figure 8:** Radial correlation function g(r) for $La_{0.5}Ca_{0.2}Sr_{0.3}MnO_3$ (x = 0.3) at 310K and 150K. Arrows indicate the two nearest neighbor bond distances for Mn sublattice.

**Figure 9:** Radial correlation function g(r) for $La_{0.5}Ca_{0.1}Sr_{0.4}MnO_3$ (x = 0.4) at 275K and 150K. Arrows indicate the three nearest neighbor bond distances for Mn sublattice.

**Figure 10:** Temperature dependence of transmitted neutron beam polarization (P) for sample $La_{0.5}Ca_{0.1}Sr_{0.4}MnO_3$ (x = 0.4) in H = 50 Oe.

**Figure 11:** Small angle neutron scattering (SANS) intensity (open circles) as a function of Q ($Å^{-1}$) for sample $La_{0.5}Ca_{0.1}Sr_{0.4}MnO_3$ (x = 0.4). The continuous line is a squared Lorentzian fit, as described in the text. The inset displays the temperature dependence of (a) correlation length ($\xi$) and (b) lorentzian amplitude $I_0$.

**Figure 12:** SANS intensity as a function of temperature at Q = 0.01 $Å^{-1}$ for $La_{0.5}Ca_{0.1}Sr_{0.4}MnO_3$ (x = 0.4) sample. The continuous line is a guide for the eye.

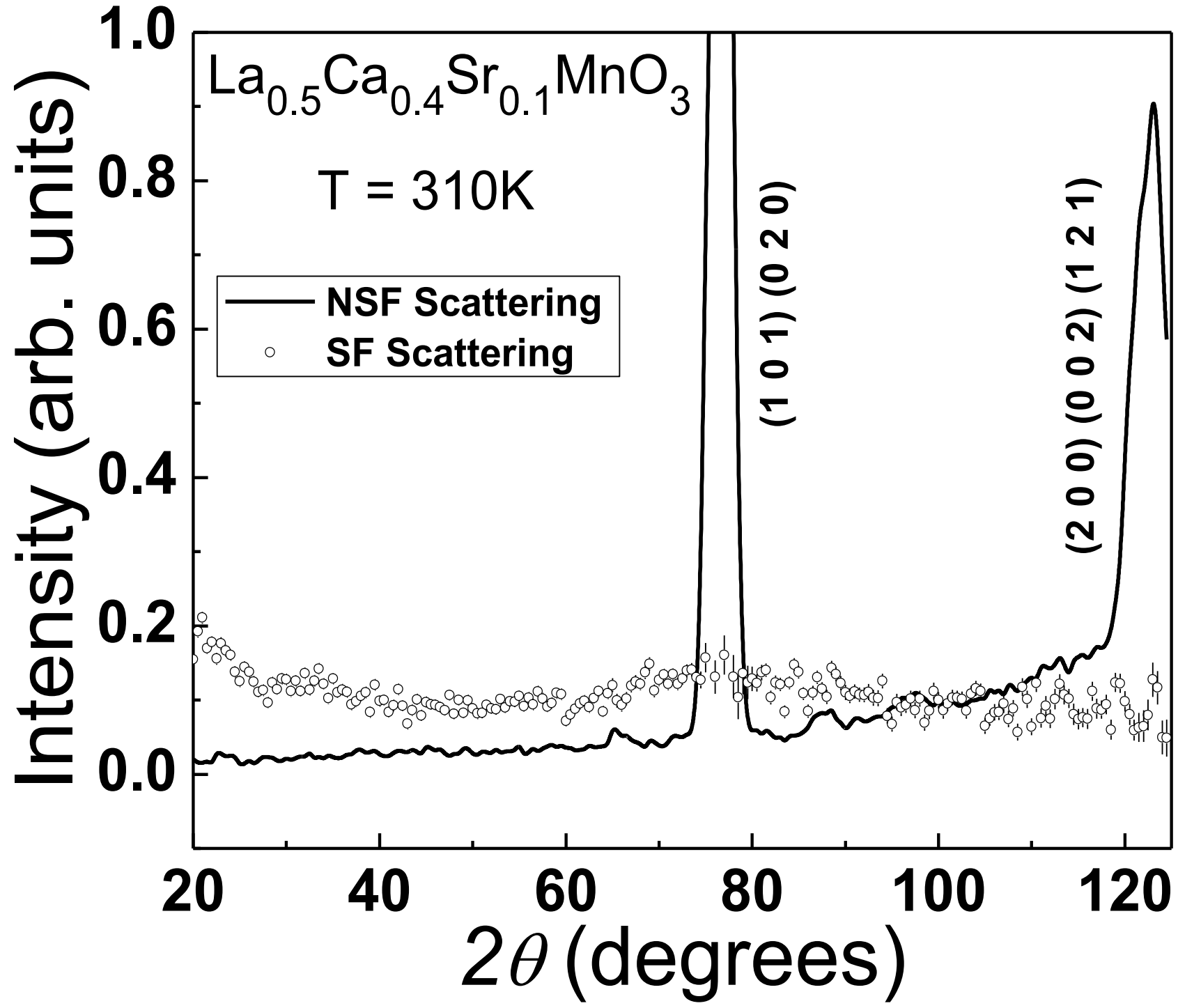

La0.5Ca0.4Sr0.1MnO3
T = 310K
NSF Scattering
SF Scattering
(1 0 1) (0 2 0)
(2 0 0) (0 0 2) (1 2 1)
Intensity (arb. units)
1.0
0.8
0.6
0.4
0.2
0.0
20
40
60
80
100
120
2θ (degrees)

Figure 1

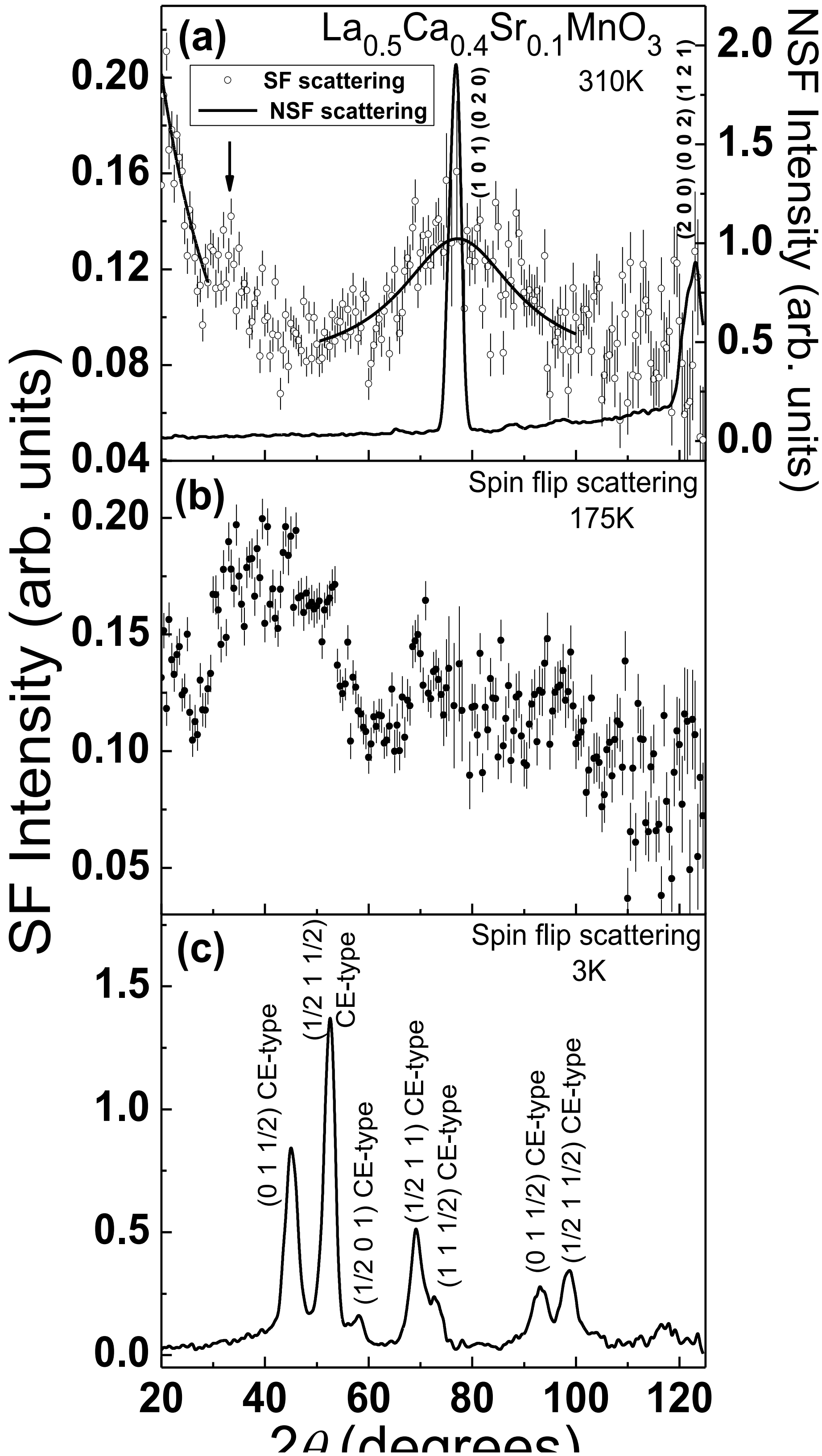
(a)
$La_{0.5}Ca_{0.4}Sr_{0.1}MnO_3$
310K
SF scattering
NSF scattering
(1 0 1) (0 2 0)
(2 0 0) (0 0 2) (1 2 1)
(b)
Spin flip scattering
175K
(c)
Spin flip scattering
3K
(0 1 1/2) CE-type
(1/2 1 1/2) CE-type
(1/2 0 1) CE-type
(1/2 1 1) CE-type
(1 1 1/2) CE-type
(0 1 1/2) CE-type
(1/2 1 1/2) CE-type
SF Intensity (arb. units)
NSF Intensity (arb. units)
2θ (degrees)

Figure 2

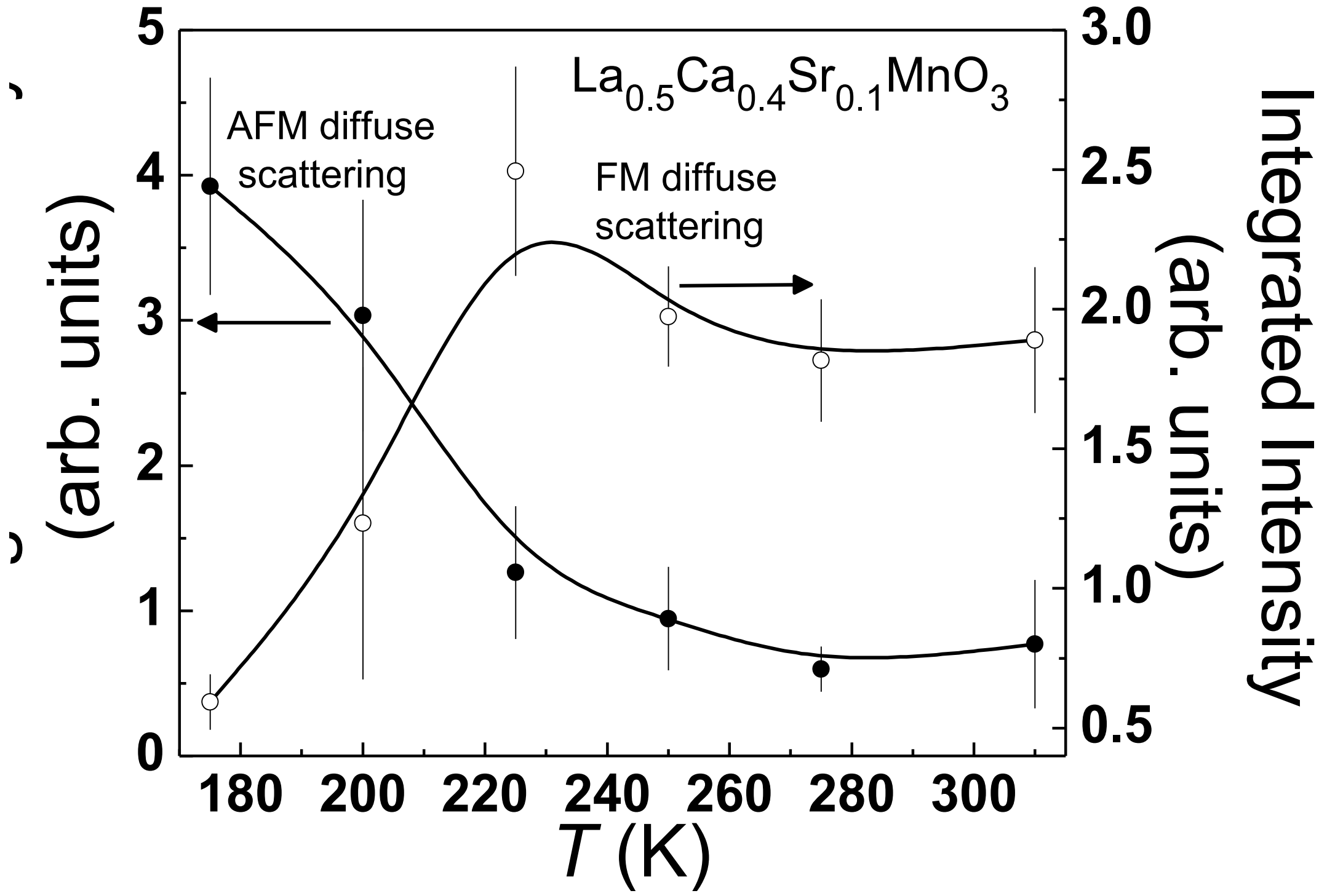

La0.5Ca0.4Sr0.1MnO3
AFM diffuse scattering
FM diffuse scattering
(arb. units)
Integrated Intensity (arb. units)
T (K)
5
4
3
2
1
0
3.0
2.5
2.0
1.5
1.0
0.5
180
200
220
240
260
280
300

Figure 3

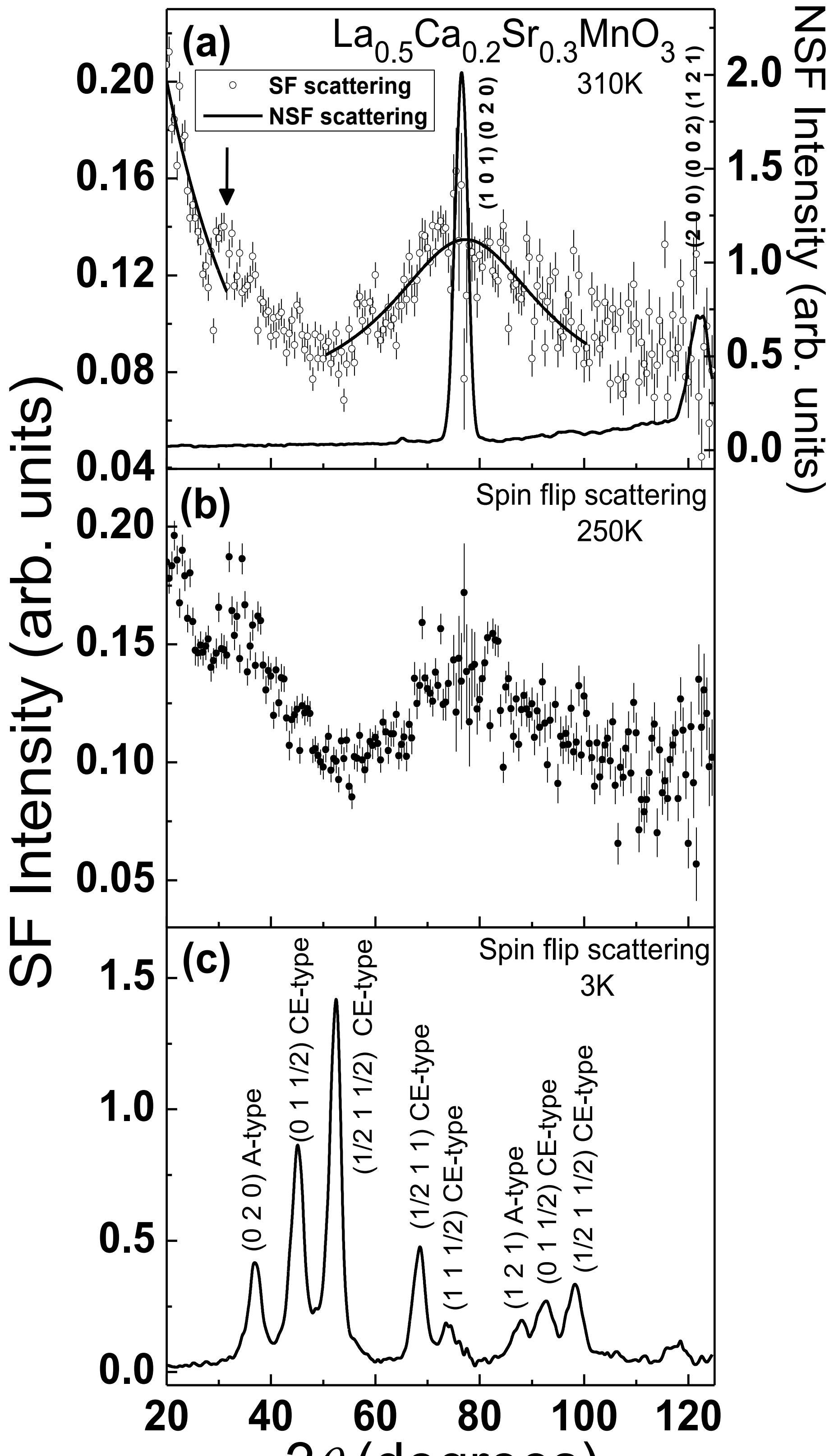

(a)
$La_{0.5}Ca_{0.2}Sr_{0.3}MnO_3$
310K
SF scattering
NSF scattering
(1 0 1) (0 2 0)
(2 0 0) (0 0 2) (1 2 1)
NSF Intensity (arb. units)
(b)
Spin flip scattering
250K
(c)
Spin flip scattering
3K
(0 2 0) A-type
(0 1 1/2) CE-type
(1/2 1 1/2) CE-type
(1/2 1 1) CE-type
(1 1 1/2) CE-type
(1 2 1) A-type
(0 1 1/2) CE-type
(1/2 1 1/2) CE-type
SF Intensity (arb. units)
2θ (degrees)

Figure 4

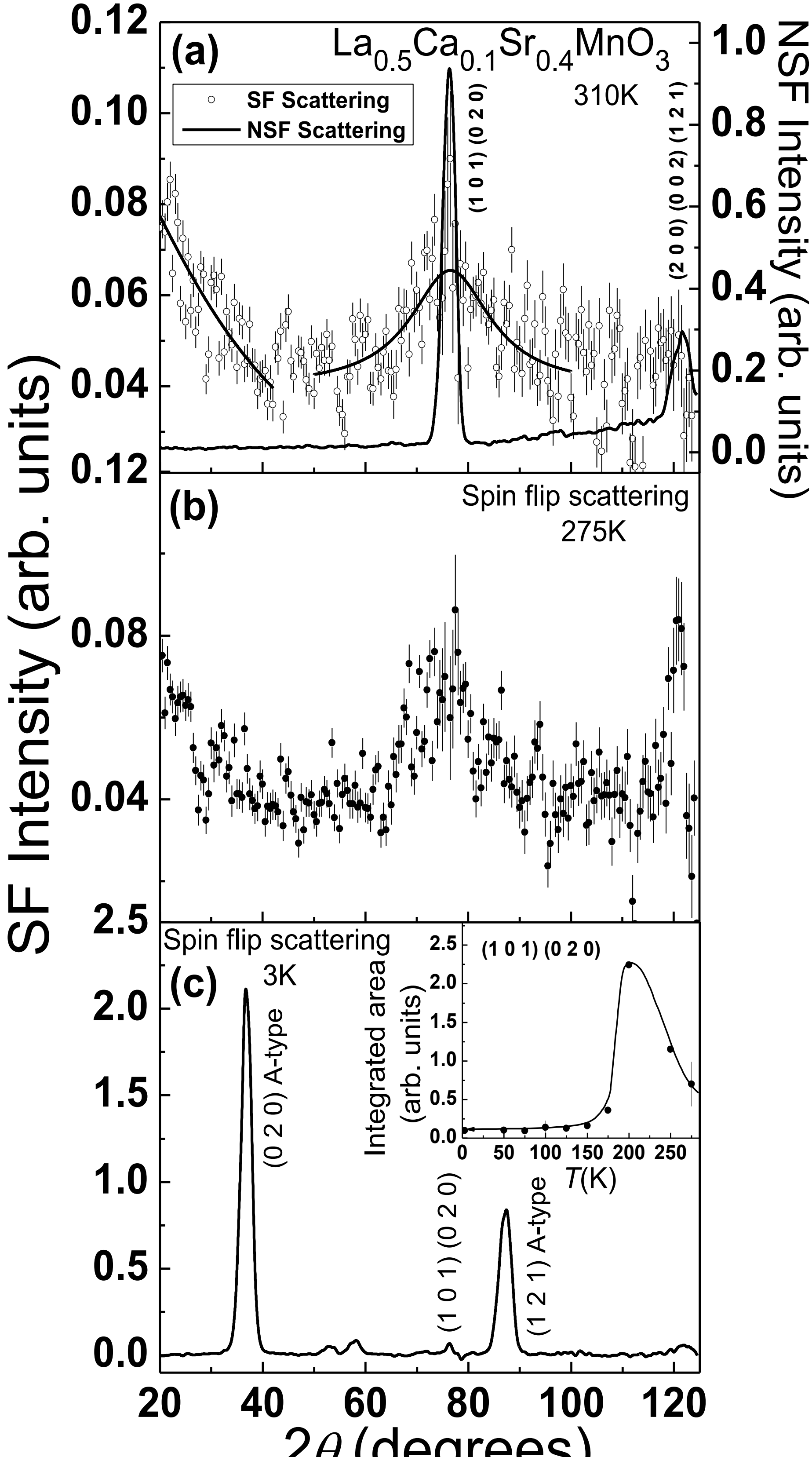
(a)
$La_{0.5}Ca_{0.1}Sr_{0.4}MnO_3$
310K
SF Scattering
NSF Scattering
(1 0 1) (0 2 0)
(2 0 0) (0 0 2) (1 2 1)
NSF Intensity (arb. units)
(b)
Spin flip scattering
275K
SF Intensity (arb. units)
(c)
Spin flip scattering
3K
(0 2 0) A-type
(1 0 1) (0 2 0)
(1 2 1) A-type
(1 0 1) (0 2 0)
Integrated area (arb. units)
T(K)
2θ (degrees)

Figure 5

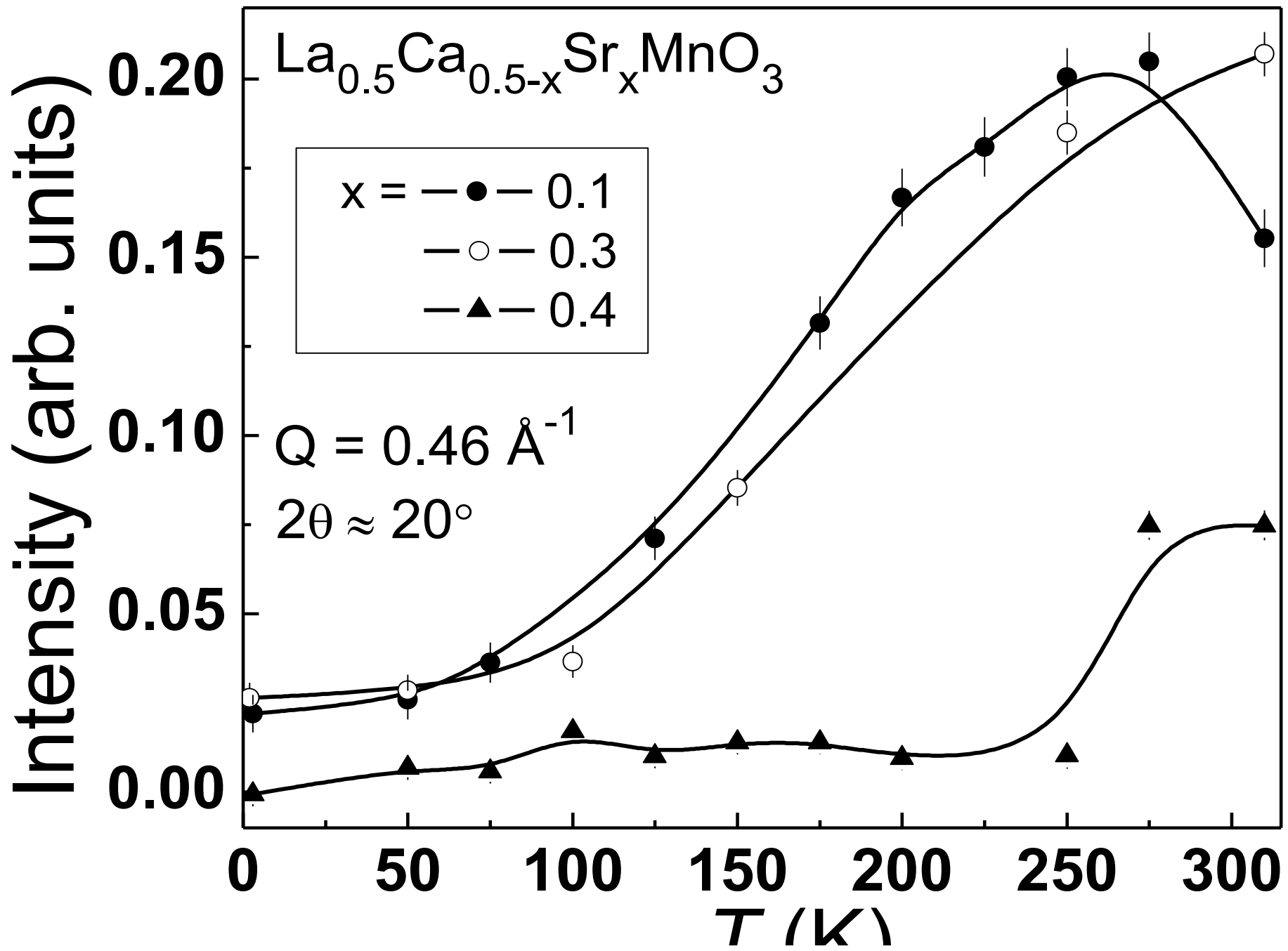

Figure 6

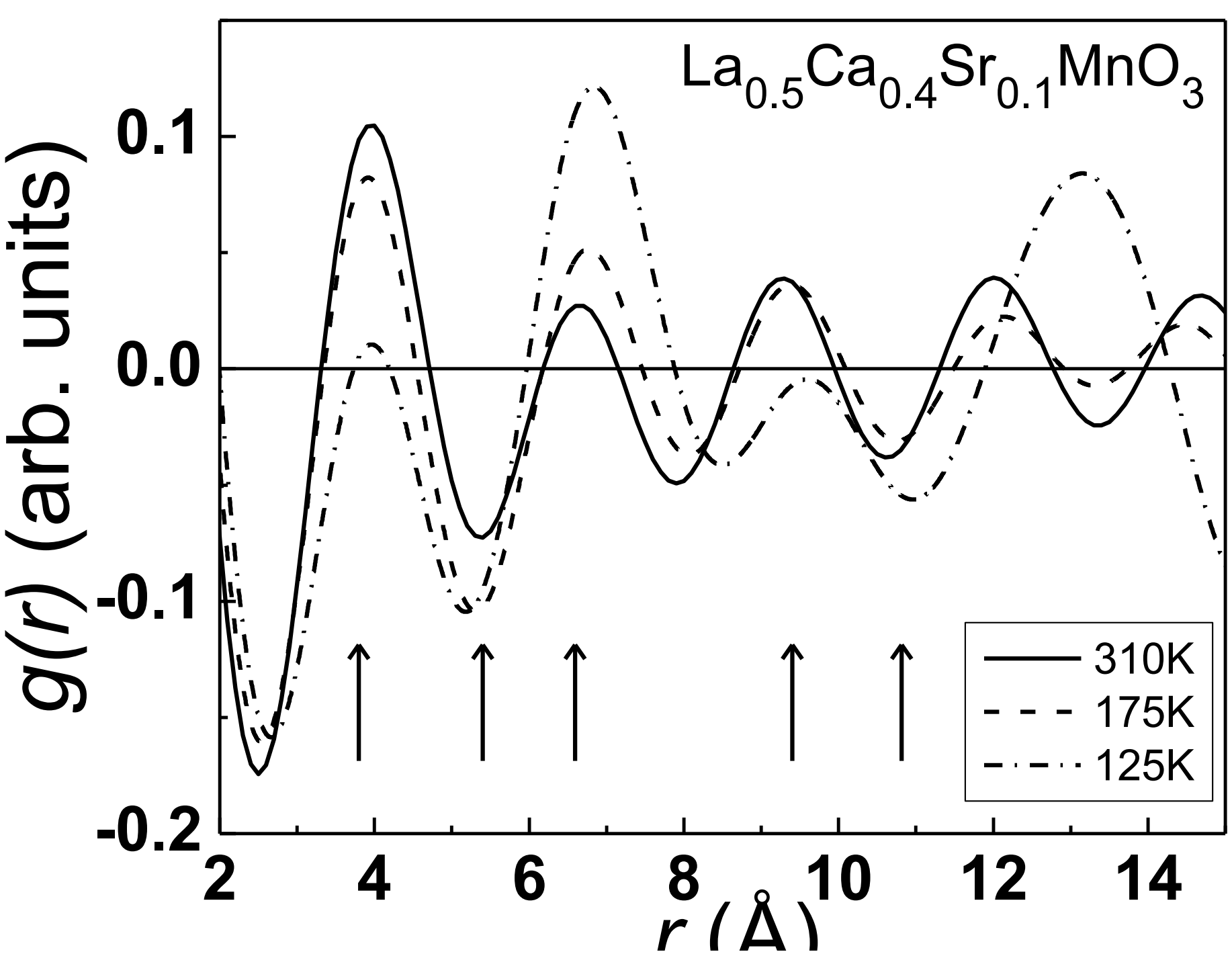

Figure 7

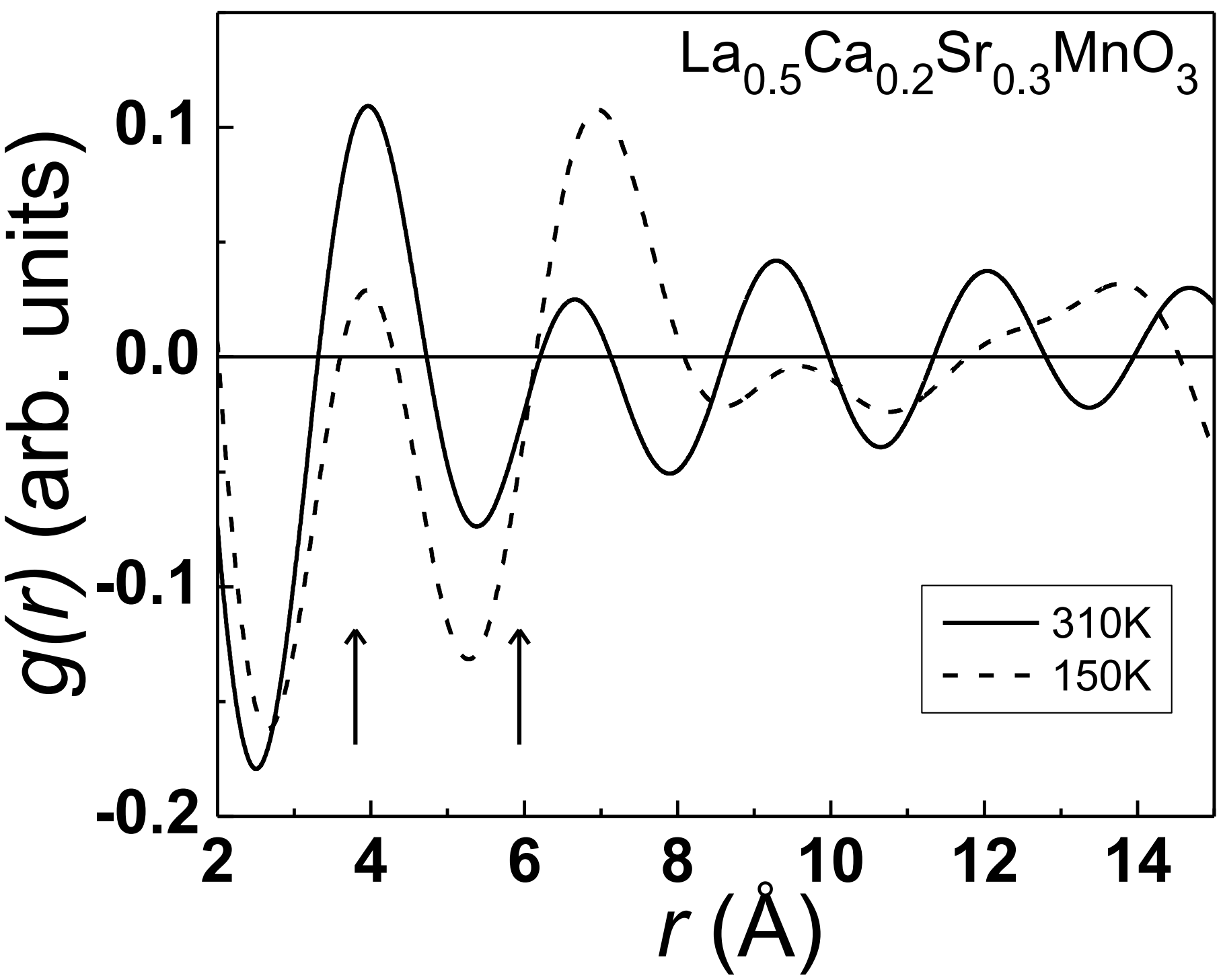

Figure 8

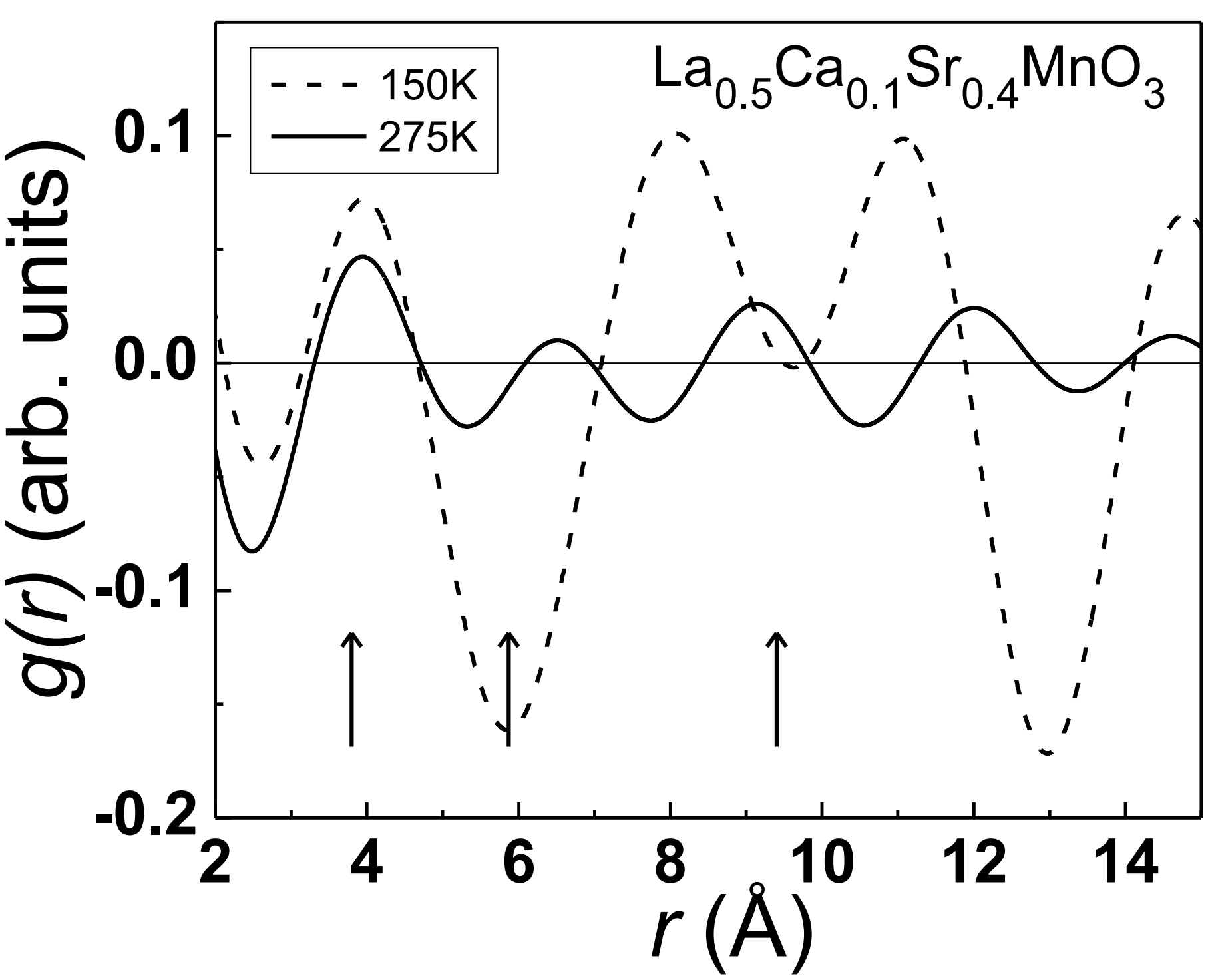

Figure 9

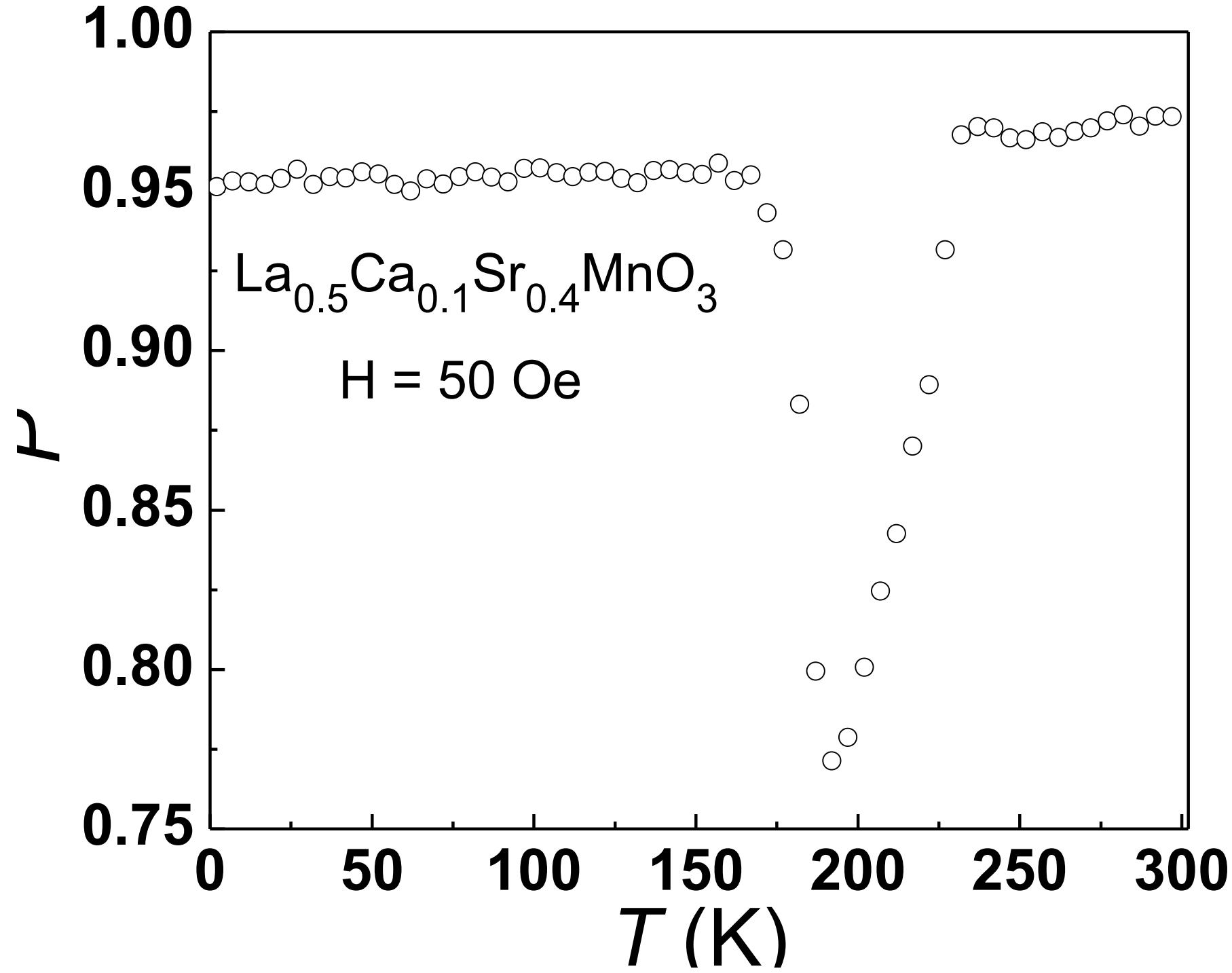

Figure 10

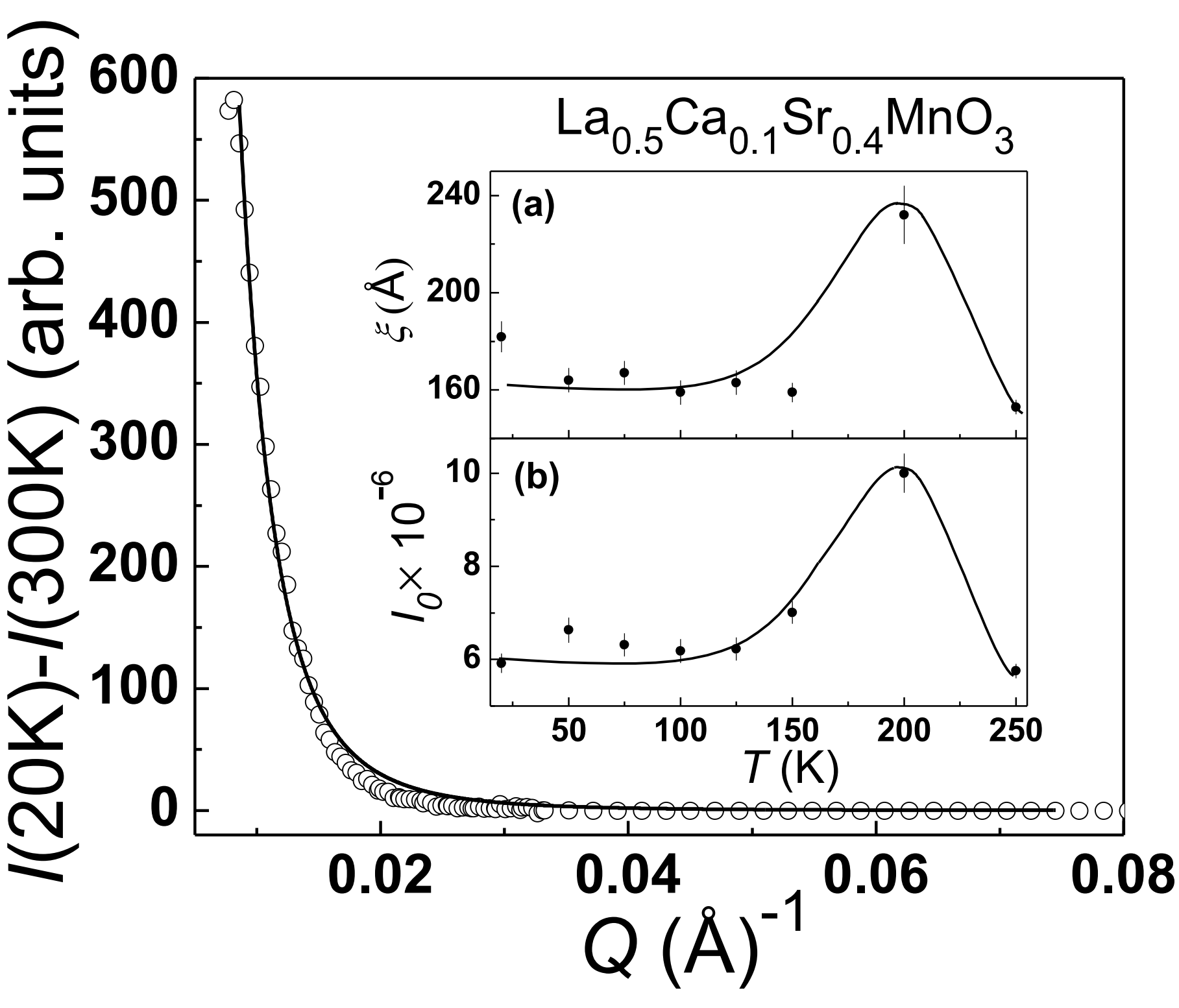

Figure 11

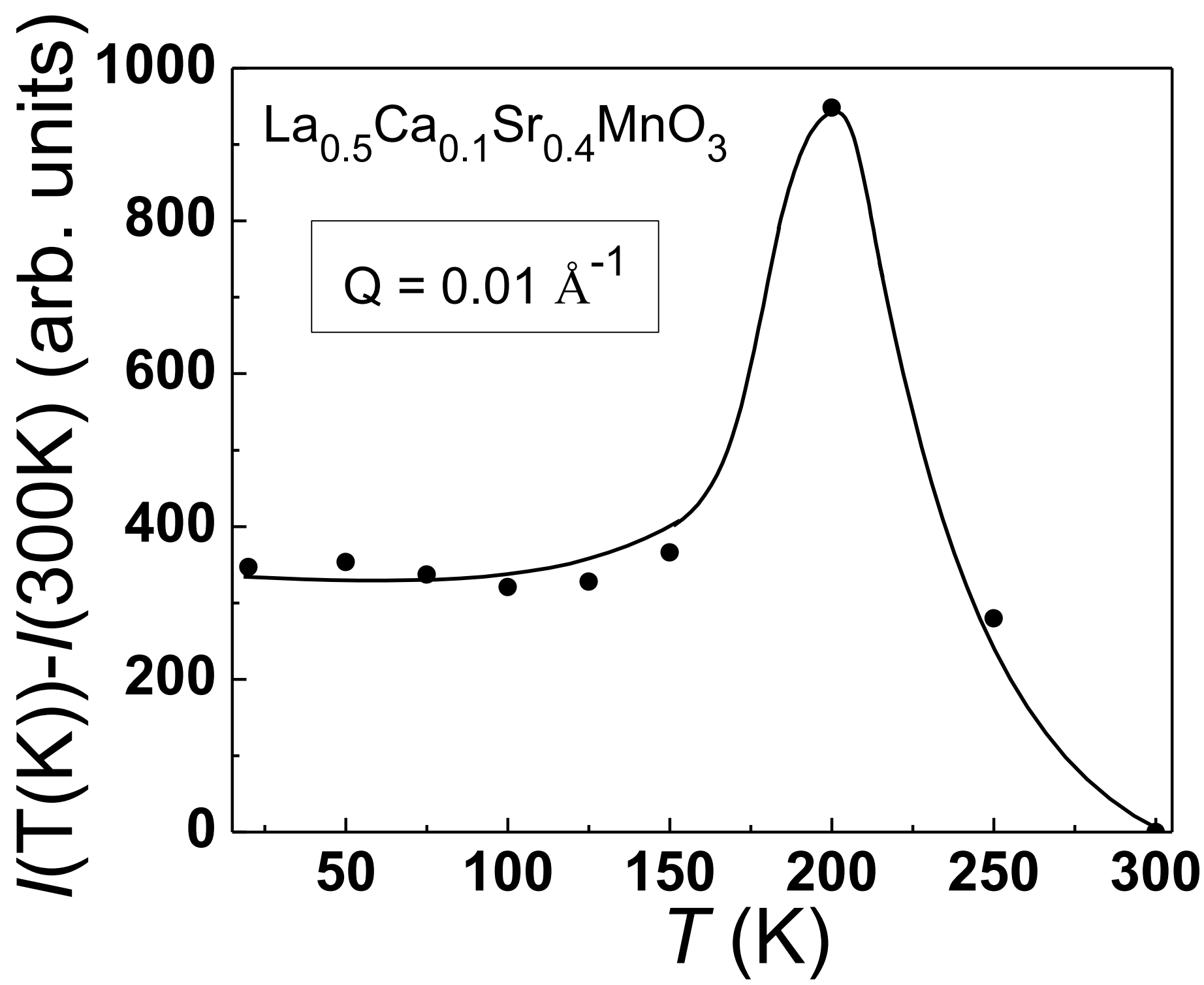
La0.5Ca0.1Sr0.4MnO3
Q = 0.01 Å-1
I(T(K))-I(300K) (arb. units)
T (K)
0
200
400
600
800
1000
50
100
150
200
250
300

Figure 12